\documentclass{ws-ijmpe}

\begin{document}

\markboth{Authors' Names}{Instructions for
Typing Manuscripts (Paper's Title)}

\catchline{}{}{}{}{}

\title{RELATIVE MOTION CORRECTION TO FISSION
  BARRIERS\\
}

\author{J. SKALSKI
}

\address{A.~So\l tan Institute for Nuclear Studies, ul. Ho\.za 69,
\\
 PL- 00 681, Warsaw, Poland \\
jskalski@fuw.edu.pl}

\maketitle

\begin{history}
\received{(received date)}
\revised{(revised date)}
\end{history}

\begin{abstract}
 We discuss the effect of kinetic energy of the relative motion
 becoming spurious for separate fragments on the selfconsistent
 mean-field fission barriers.
 The treatment of the relative motion in the cluster model
 is contrasted with the necessity of a simpler and approximate
 approach in the mean-field theory. A scheme of the energy
  correction to the Hartree-Fock is proposed.
  The results obtained with the effective Skyrme interaction SLy6
  show that the correction, previously estimated as $\sim$ 8 MeV in $A=70-100$
  nuclei, amounts to 4 MeV in the medium heavy nucleus $^{198}$Hg and
  to null in $^{238}$U. However, the corrected barrier
  implies a shorter fission half-life of the latter nucleus.
  The same effect is expected to lower barriers for multipartition (i.e.
  ternary fission, etc)
  and make hyperdeformed minima less stable.
\end{abstract}

\section{ Introduction}

 It seems that the existing calculations of fission barriers overestimate
 energies of nuclear configurations close to scission by including a
 spurious contribution of kinetic energy of the fragments' {\it relative}
  motion.
 This may be seen as follows: The binding of two separated selfbound systems
  is equal to $E_{sep}=B_1+B_2+V_{int}$, with $B_i$ fragment binding energies
  and $V_{int}$ Coulomb energy.
 For a compound system, energy $E(1+2)$, should tend to $E_{sep}$
 for separate entities, hence it should not contain fragment center
 of mass (c.m.) energies $E_{c.m.}(i)$. However, in a standard Hartree-Fock
 (HF), energy of a two-piece configuration still contains the term
 $E_{c.m.}(rel)=E_{c.m.}(1)+E_{c.m.}(2)-E_{c.m.}(1+2)$,
 corresponding to the relative motion of two fragments \cite{BG80,JSnew}.

  Within the mean-field theory, this overestimate arises from the
  c.m. kinetic energy correction that has to be subtracted from
 the expectation value of the Hamiltonian (energy functional) to
  obtain the binding. The expectation value
 of the operator $(\sum_k {\bf p}_k)^2/(2AM)$ on the Slater state reads:
  \begin{equation}
  \label{HFK}
  E_{c.m.}(A)=\frac{1}{2AM}\left(\sum_{k=1}^A \langle k\mid{\bf p}^2\mid k\rangle
  -\sum_{k\ne l}^A \mid\langle k\mid{\bf p}\mid l\rangle\mid^2\right),
 \end{equation}
 with $k,l$ labelling occupied single particle states, and $M$ nucleon mass.
  One should include one $E_{c.m.}$ correction for a compound system, two
  corrections for two separate fragments, three corrections for three
  fragments, etc.
  For two distant fragments, $A_1+A_2=A$, if one could {\it distinguish}
  particles belonging to each, which would imply the vanishing expectation
 value $\langle(\sum_{k\in 1}{\bf p}_k)\cdot(\sum_{l\in 2}{\bf p}_l)\rangle$,
  \begin{equation}
  \label{as}
  E_{c.m.}(A_1)+E_{c.m.}(A_2)=E_{c.m.}(A)+\frac{A_2 E_{c.m.}(A_1)+A_1
 E_{c.m.}(A_2)}{A} .
  \end{equation}
 The second term on the right-hand side is just the asymptotic (for
 the distance $R\rightarrow\infty$) value of
 kinetic energy of the fragments' relative motion, $E_{c.m.}(rel)=
   <{\bf P}^2_{rel}>/(2\mu)$, with ${\bf P}_{rel}=
 (A_2/A)\sum_{k\in 1}{\bf p}_k-(A_1/A)\sum_{k\in 2}{\bf p}_k$,
 and $\mu=MA_1A_2/A$. This quantity becomes spurious for a two-piece system
  as it does not contribute to the binding.
 As noted by Berger and Gogny, \cite{BG80} this asymptotic term should
 be subtracted from the HF energy to obtain proper fusion barriers.

  In practical HF calculations the $E_{c.m.}$ correction is included in
 various forms (see \cite{BRRM}). Here we confine the discussion to
 effective interactions which use the
 natural definition (\ref{HFK}). With such interactions one obtains
  $E_{c.m.}=5.5 - 8$ MeV for $A=40-250$, decreasing with $A$.
  This quantity should be subtracted
  from $E(1+2)$ in a consistent theory:
 partially - for shapes with constriction, totally - for two-piece systems.
 This subtraction is usually included in calculations of fusion barriers
 (otherwise the barriers are too high \cite{DN},\cite{Dobr},\cite{JSn}),
  but omitted in fission studies.
 However, even for fusion barriers, a gradual dependence of the correction
 on the compactness of the system is missing.

 As found in Ref. \cite{JSnew}, the subtraction of the asymptotic value of
 $E_{c.m.}(rel)$ brings the calculated HF fission barriers closer
 to the experimental values in medium-size $A= 70-100$ nuclei.
 In the present work we estimate the effect of the shape-dependent correction
  $E_{c.m.}(rel,shape)$ on fission barriers in heavier nuclei, using
   the Skyrme effective interaction SLy6  \cite{SLy6} (section 3).
 The correction is discussed and defined in section 2, where we also
  consider a different treatment of the relative kinetic energy in the
 cluster model and in the mean-field theory.
 Conclusions are given in section 4.

 It is remarkable that a correction of similar property and magnitude,
 although based on completely different grounds, \cite{MSC}
 has been introduced in macroscopic-microscopic calculations
  \cite{fiss,newfis} in order to obtain a better agreement with
  the experimenal fission barriers in the same regions of nuclei.

\section{General discussion}

 The difficulty in the determination of the fission or fusion barrier discussed
 here is pertinent to nuclear models in which the binding energy of the
  far separated nuclei 1 and 2 treated as a one system is different from
 $B_1+B_2$, with $B_i$ determined separately. Since the standard HF
 belongs to this category, it needs a correction which would ensure
  that $E(1+2)\rightarrow B_1+B_2$ in fission (with $E(1+2)$ understood
 as adiabatic energy). The smoothness of such correction with the
 evolving nuclear shape is a natural requirement.

  At the heart of the difficulty lies the identity of particles that
  impedes a definition of the relative coordinate and motion
  of two interacting subsystems. This problem is crucial in studies of
  light nuclei, where it is treated within the resonatig group method
  (RGM) that is basically a version of the generator coordinate method
  (GCM). A cluster configuration $A_1+A_2$ is
  represented as an expansion
  $\Phi_A=\int d{\bf r} \varphi({\bf r})\Phi_{\bf r}$
  onto an overcomplete basis formed by states
   $\Phi_{\bf r}={\cal A}_{A_1 A_2} [\delta({\bf r}-{\bf r}_{rel})\Phi_{A_1}
   \Phi_{A_2}]$ \cite{Lov98},\cite{SLYV03} with
$\Phi_{A_i}$ the cluster states depending on {\it intrinsic}
  coordinates, $\varphi$ the amplitude of the relative motion and
 ${\cal A}_{A_1 A_2}$ the antisymmetrizer containing permutations
  mixing the coordinates of the first $A_1$ with those of
  the last $A_2$ nucleons. Thus, the label ${\bf r}$ of the basis states
  assumes the role of the intercluster coordinate.

  From the Schr\"odinger equation for $\Phi_A$ one obtains the Hill-Wheeler
  equation for the amplitude of the relative motion in coordinate ${\bf r}$,
  with a well defined Hamiltonian. However, a decomposition of this
  Hamiltonian into kinetic and potential parts, ${\cal T}+{\cal U}$, is
  arbitrary: The relative kinetic energy operator is {\it assumed}
 as ${\cal T}=-(\hbar\nabla_{\bf r})^2/(2\mu)$ with the reduced mass
 $\mu$, and this {\it fixes} the potential
  ${\cal U}$ \cite{Lov98},\cite{SLYV03}.
 So obtained potentials are much deeper in the compound nucleus
 region than those implied by the mean field; in this way the Pauli exclusion
 influences the relative motion of the overlaping clusters.

 It follows that while the RGM (or GCM) provides
 a solution to the relative motion problem, its ingredients, like the potential
  ${\cal U}$, do not seem to be of much use for the mean-field theory.
 As an aplication of the full RGM (GCM) method for heavy nuclei would be
  prohibitively difficult, one would rather include a kinetic energy
  correction in the relatively simple HF method to improve energy asymptotics
  for separated clusters.
  However, this cannot be just the expectation value of
  $[A/(2MA_1A_2)]{\bf P}^2_{rel}$ in the Slater state:
  Owing to the incompatibility of the ${\bf P}_{rel}$ variable with the
  antisymmetry of the Slater determinant, its value, $(<{\bf p}^2>
 +\sum_{k\ne l}^A \mid\langle k\mid{\bf p}\mid l\rangle\mid^2/[A(A-1)])/(2M)$,
  is by $\sim$ 10 MeV larger than the proper value of $E_{c.m.}(rel)$.
  This is why obtaining the correct value of Eq.(\ref{as}) requires an
  extention that goes beyond HF. Some guidance might be provided by
  realistic internuclear potentials used in fusion studies, e.g. \cite{SL79}.
  These potentials have correct two-fragments asymptotics, but should be
  smoothly replaced by the mean-field energy for compact mononuclear
  configurations, where they become irrelevant.

  One possibility to proceed \cite{JSn} is to introduce a measure
 of the fragment separation $\xi$ which would replace the relative distance
 ${\bf r}$ and define the subtracted portion
 of the relative kinetic energy, $E_{c.m.}(rel,shape)=\xi E_{c.m.}(rel)$.
  To this aim, consider dispersion of the number of particles in the $k$-th
 HF orbital, residing in the volume $V_1$ of the first fragment with $A_1$
 nucleons. This reads $p_k(1-p_k)$, with $p_k=\int_{V_1}\mid \psi_k \mid ^2$
  (with many-particle correlations ignored).
 For completely divided fragments the $k$-th wave function is localized, so
 $p_k=$0 or 1 and dispersion vanishes. We define $\xi$ by means of
  dispersion averaged over the occupied orbitals:
  \begin{equation}
  \label{xi}
 \xi=1-\left(\frac{2}{A}\right)\frac{\sum_{k-occ}p_k(1-p_k)}{{\bar p}(1-{\bar p})}=
 \left(\frac{2}{A}\right)\frac{\sum_{k-occ}(p_k-{\bar p})^2}{{\bar p}(1-{\bar p})}  ,
  \end{equation}
 with ${\bar p}=2\sum_{k-occ} p_k/A=A_1/A$ (with obvious modifications for
 pairing included). So defined $\xi$ is positive,
 reaches the maximal value 1 for separated fragments and falls to
 zero for wave functions uniformly smeared over two fragments.
 It has thus necessary properties to show main effects of a gradual
  inclusion of $E_{c.m.}(rel)$ in HF energy.

 Subtraction of a varying portion of $E_{c.m.}(rel)$ will change
 energy balance between configurations with and without constriction,
 lowering the former with respect to the latter. As found in the study
 \cite{JSn} of fusion barriers, values of $\xi$ at the barrier
 vary between 0.7 and 0.9 and decrease together with the interfragment
 distance $R$. The latter is defined as the distance between the centers of
 mass of two half-spaces, containing $A_1$ and $A_2$ nucleons.
 Clearly, fusion barriers calculated using the shape-dependent correction
 $\xi E_{c.m.}(rel)$ are higher than those obtained by subtracting the whole
  asymptotic value $E_{c.m.}(rel)$.
  The related increase in the barrier height will depend
  on the slope $dV/dR$ (without any correction): a small increase
  for a large positive slope, a larger increase
  (and the barrier shift towards smaller $R$) for small positive or negative
  slopes. For example, with the SLy6 force,
 the inclusion of the $\xi$-dependence leads to
   the increase in fusion barrier by 1.8 MeV for $^{48}$Ca$+^{48}$Ca and
  by 2.5 MeV MeV for $^{48}$Ca$+^{208}$Pb, with the inward shift of the
   barrier top by 0.6 and 1.5 fm, respectively \cite{JSn}.

  In the present study of fission barriers we use a different prescription
  for the relative kinetic energy correction.
  For a system of $A$ nucleons consider its division $A_1+A_2$ into volumes
  $V_1$ and $V_2$. Calculate the quantity $p_k$ for
  each wave function and call it localized in $V_1$ ($V_2$) if $p_k>
  1-\epsilon$ ($p_k< \epsilon$), with some small $\epsilon$ (we use
  $\epsilon=0.03$). Suppose that for a given nuclear shape (configuration)
   $N_1$ wave functions are localized in $V_1$ and $N_2$
  in $V_2$. Then the correction for this shape is defined as
  \begin{equation}
 \label{corr}
   E_{c.m.}(rel,shape)=
  \frac{N_1}{A_1}E_{c.m.}(N_1)+\frac{N_2}{A_2}E_{c.m.}(N_2)-\frac{N_1+N_2}{A}
  E_{c.m.}(N_1+N_2) .
  \end{equation}
  This quantity tends to $0$ for no localization and to the asymptotic value
  $E_{c.m.}(rel)$ for divided fragments (full localization). The
  correction (\ref{corr}) is more directly related to the localized
  orbitals than $\xi E_{c.m.}(rel)$, but still not completely satisfactory.
    Ultimately, it would be desirable
  to define the correction for relative kinetic energy as a part of the
   energy functional and treat it variationally.

\section{Shape-dependent correction to barriers}

  We have calculated the fission barriers with and
  without the $E_{c.m.}(rel,shape)$ correction in $^{198}$Hg and $^{238}$U.
  Pairing was included as the delta interaction in the BCS scheme, using
   the cutof according to the prescription of Ref. \cite{cutof}.
   The delta interaction strength
  was fixed at $V_n=316$ MeV fm$^3$ for neutrons and $V_p=322$ MeV fm$^3$ for
  protons.
\begin{figure}[t]
\hspace{30mm}
\begin{minipage}[t]{50mm}
\centerline{\includegraphics[scale=0.3, angle=-90]{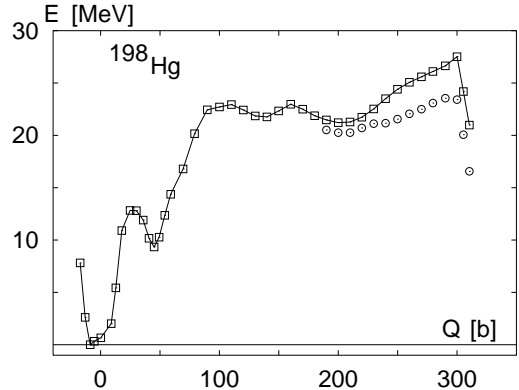}}
\end{minipage}
\caption{{\protect Fission barrier in $^{198}$Hg
 without (squares) and with (circles) the $E_{c.m.}(rel,shape)$ correction.
  }}
\end{figure}

  The calculated fission barrier in $^{198}$Hg (Fig. 1) is mass-symmetric
  ($A_1=A_2$) and the saddle corresponds to a large elongation with
  the quadrupole moment close to $Q=300$ b, cf Fig. 2.
  To relate Fig. 1 to other studies, we
  mention that the energy plot for $Q< 20$ b, calculated with the Gogny
  interaction, may be found in Fig. 2 of \cite{Eg96}, while the whole
  macroscopic-microscopic fission barrier was given in Fig. 3 of \cite{Mos72}.
    Between the secondary minimum at $Q=45$ b and
  $Q\approx 100$ b the HF minima are soft to mass-asymmetry or even slightly
  mass-asymmetric.
  Our calculated barrier of 27.5 MeV is lowered owing to the
  $E_{c.m.}(rel,shape)$ correction to
  23.7 MeV  vs. the experimental value of 20.4 MeV \cite{Mor72} and 19.3 MeV
  calculated in \cite{Mos72}. It may be seen that the
  relative kinetic energy correction of 3.8 MeV at the $^{198}$Hg fission
  barrier is smaller than those for $A=70-100$ nuclei \cite{JSnew}, but still
  significant.
\begin{figure}[t]
\hspace{40mm}
\begin{minipage}[t]{30mm}
\centerline{\includegraphics[scale=0.3, angle=0]{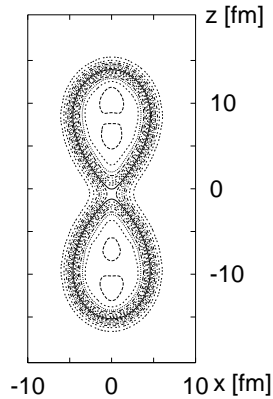}}
\end{minipage}
\caption{{\protect Barrier configuration in $^{198}$Hg, $Q=$300 b;
  density contours 0.01 fm$^{-3}$ apart.
  }}
\end{figure}

  The calculated barrier in $^{238}$U is shown in Fig. 3. The triaxial
  first hump of 7.55 MeV at $Q=60$ b and the mass-asymmetric second hump
  of 8 MeV at $Q=125$ b (that agrees with the result with SLy6 reported
  in \cite{B}) are both larger than the experimental barriers
  (inner and outer) close to 6 MeV \cite{BL80},\cite{Smir}. At
 the second barrier, the neck is still not yet developed (cf shapes in Fig. 4)
 and single-particle orbitals are not well localized in
  parts of the volume corresponding to $A_1=138$ and $A_2=100$, the
  partition chosen close to the maximum of the experimental mass yield
  \cite{Wahl}. Hence the $E_{c.m.}(rel,shape)$ correction vanishes and
  does not influence the barrier height in this case.
  For quadrupole moments $Q>125$ b the localization
  begins and the correction grows slowly with $Q$. At $Q=200$ b (shape in
  Fig. 4) it amounts already to 1.9 MeV. This implies that the barrier
  relevant for the quantum tunneling, while not being higher, becomes shorter.

  An estimate of the effect on the fission half-life may be obtained by
  using the relation $\Delta \log T_{sf}\approx 0.8686 \Delta S$, with action
   $S=\int \sqrt{2B_{eff}(E-E_{g.s})}d\beta_2$, see e.g. \cite{sss95}.
   For $B_{eff}$ we can use the cranking mass parameters, typical for the
  appropriate deformation range, calculated with the Woods-Saxon potential.
  An approximate relation between the quadrupole moment and the deformation
  parameter $\beta_2$, $\beta_2\approx \sqrt{5\pi}Q/(3r_0^2A^{5/3})$, with
  the assumed $r_0=1.2$ fm, gives $\beta_2\approx 0.01\times Q$ for $A=238$.
  Between $\beta_2=1.8$ and 2.3, where the correction causes an appreciable
  difference in the integrand of action, the mass parameter $B_{2 2}$
  decreases from about 20 to 5 $\hbar^2$/MeV and is similar to $B_{2 3}$,
  while $B_{3 3}$ increases from about 30 to 60 $\hbar^2$/MeV (subscripts of
  the mass tensor refer to deformation parameters $\beta_{\lambda}$,
  \cite{sss95}).
  Taking $B_{eff}=10 \hbar^2$/MeV (as $\beta_3$ increases with $\beta_2$,
  $B_{eff}>B_{2 2}$), and estimating $\Delta S$ as
  $(2.3-1.8)\times\sqrt{2B_{eff}E_{av}}$ with $E_{av}$ equal to 1.0 MeV
  (cf Fig.3),
  we obtain a rough estimate $\Delta S=2.23$ and $\Delta \log T_{sf}=1.94$.
  Thus, the expected change in the fission half-life is about two orders of
  magnitude, compared to the experimental value of $\log T_{sf}$ of 23.4
  \cite{BL80}. The true correction $\Delta \log T_{sf}$ is probably at
  least that large: The recently calculated selfconsistent
  cranking inertia parameters \cite{BSDN07}, that should be used in the exact
  calculation, seem larger than the Woods-Saxon cranking mass parameters.

  It is worth mentioning that the calculated barriers will be still lowered by
  the rotational correction. In $^{238}$U, one can expect more than a 1 MeV
  correction at the first barrier, and more than a 2 MeV correction
  at the second barrier, based on calculations \cite{Bon04}.
  An even larger rotational correction should be expected
  for $^{198}$Hg at the barrier, owing to a larger deformation.
\begin{figure}[t]
\hspace{30mm}
\begin{minipage}[t]{70mm}
\centerline{\includegraphics[scale=0.3, angle=-90]{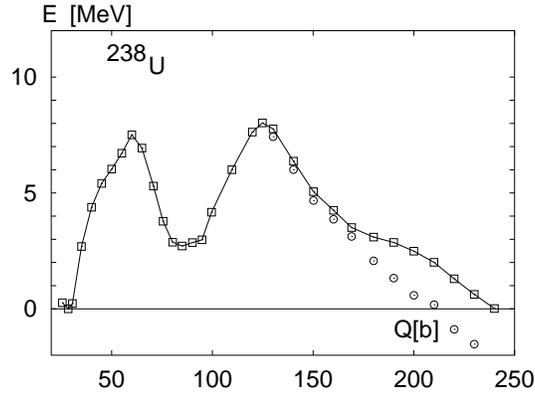}}
\end{minipage}
\caption{{\protect Fission barrier in $^{238}$U
 without (squares) and with (circles) the $E_{c.m.}(rel,shape)$ correction.
  }}
\end{figure}
\begin{figure}[t]
 \hspace{30mm}
\begin{minipage}[t]{30mm}
\centerline{\includegraphics[scale=0.3, angle=0]{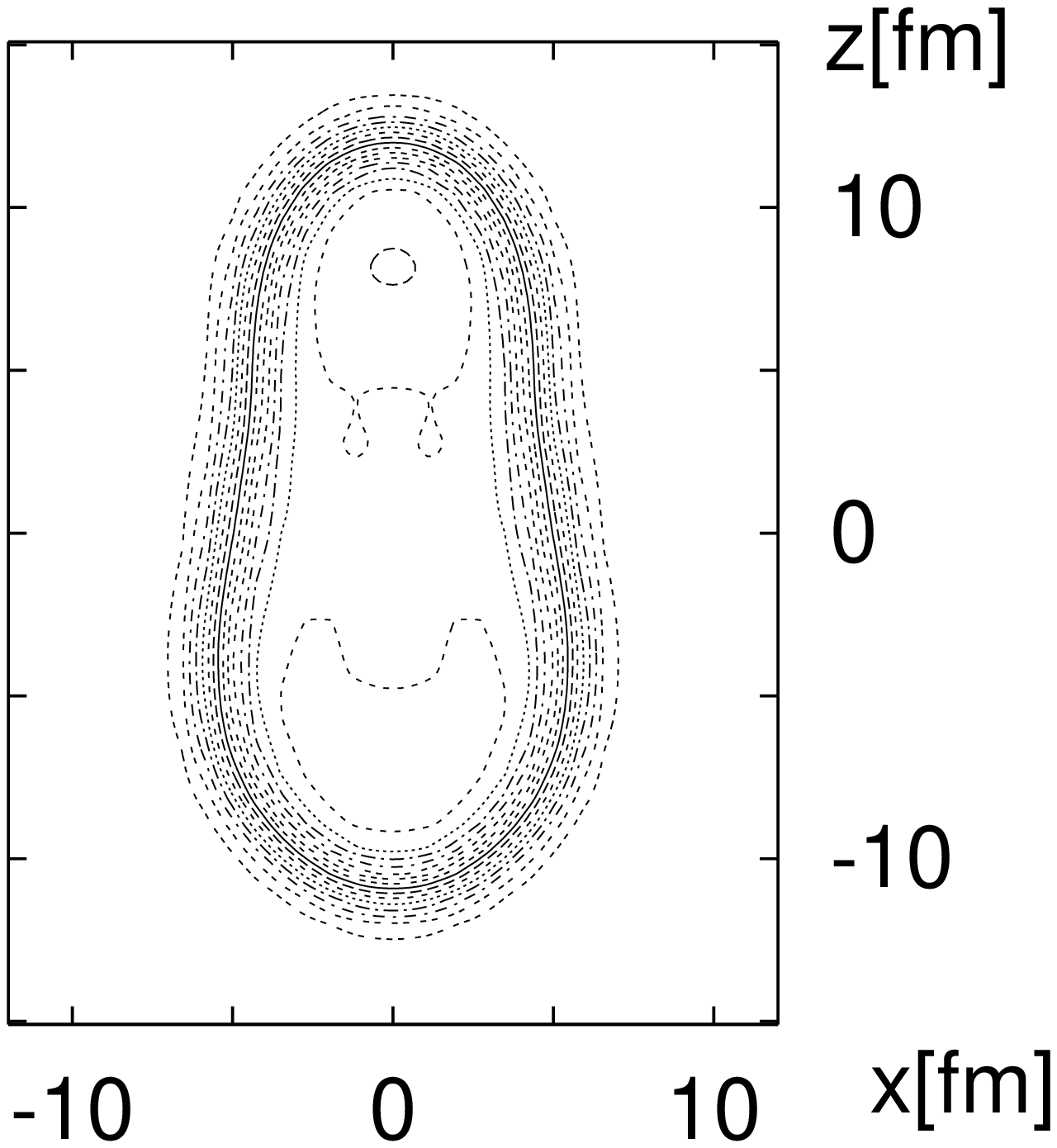}}
\end{minipage}
\hspace{10mm}
\begin{minipage}[t]{30mm}
\centerline{\includegraphics[scale=0.3, angle=0]{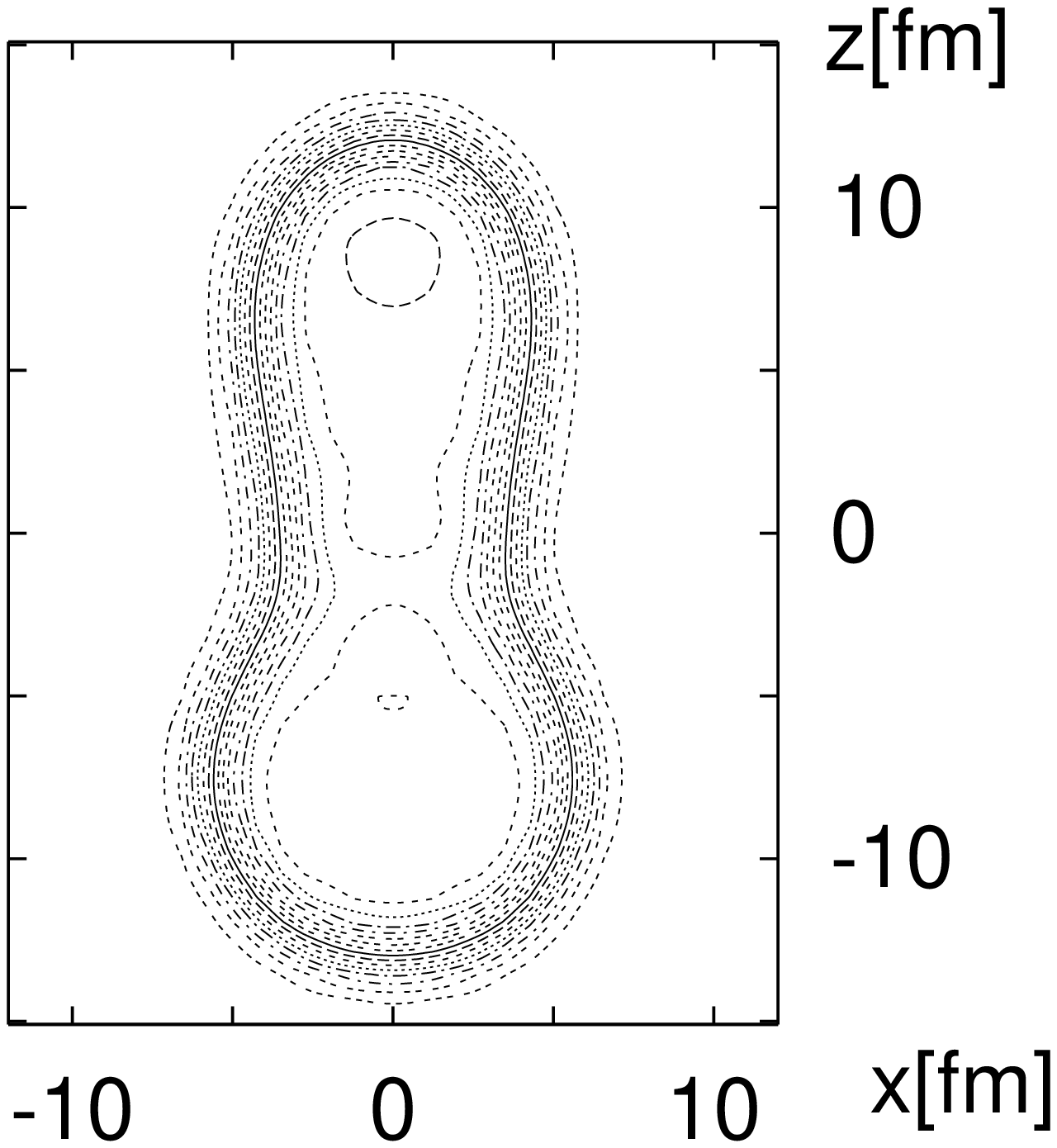}}
\end{minipage}
\caption{{\protect The barrier ($Q=$125 b, left) and beyond the barrier
 ($Q=$200 b, right) configurations in $^{238}$U;
 density contours 0.01 fm$^{-3}$ apart.
  }}
\end{figure}

\section{Conclusions}

 Correctly calculated energies of nuclear shapes with constriction
 become lower than in the standard approach, so such shapes,
  in particular scission configurations, are less excited with respect to
  more compact configurations.
 Here are some consequences of this correction for theoretical predictions:
\begin{romanlist}[(ii)]

 \item Fission barriers with configurations close to scission
     are lowered by aproximately $E_{c.m.}$ which weakly depends on the mass
     number $A$. This brings the calculated fission barriers in relatively
  light ($A=70-100)$ nuclei much closer to the data \cite{JSnew}.

 \item A smaller correction is expected for fission barriers in $A\approx 200$
   nuclei, e.g. in $^{198}$Hg (Fig. 1).
  This is consistent with very elongated, constricted shapes at the
  fission saddles in these nuclei, quite different, however,
  from scission configurations.

 \item Fission barriers with energy at the scission point close
  (within a few MeV) to that at the ground state become shorter, and this
  leads to a moderate decrease in fission half-lives (Fig. 3 for $^{238}$U).

\end{romanlist}

  Modifications are expected for barriers and half-lives for
    multipartition and multifragmentation; a scission configuration for
  triparition will be lowered by $\sim 2 E_{c.m.}$, that for the decay into
  four fragments by $\sim 3 E_{c.m.}$, etc. One can also notice that
  the correction to barriers would tend to destabilize hypothethic
  hyperdeformed minima studied in Ref. \cite{Hyp07}

  Due to the magnitude of the correction, it lowers substantially
  fission barriers (up to the actinides) and modifies fission life-times,
   except for the superheavy nuclei.
  Even there, the correction should be accounted for when considering
  fission dynamics.
 A related correction may be necessary in methods other than HF, unless
 they correctly and smoothly describe binding during fission and fusion.

\section*{Acknowledgements}
This work was supported in part by the Polish Committee for Scientific Reserch
 (KBN) Grant No. 1P03B06427 and the Polish Ministry of Science;
 by the National Nuclear Security
Administration under the Stewardship Science Academic Alliances
program through the U.S. Department of Energy Research Grant
DE-FG03-03NA00083; by the U.S. Department of Energy under Contract
No. DE-AC05-00OR22725 with UT-Battelle, LLC (Oak Ridge National
Laboratory).

 \end{document}